\documentclass[mathematics,article,moreauthors,pdftex]{my-mdpi} 

\firstpage{1} 
\pubvolume{xx}
\issuenum{1}
\articlenumber{5}
\pubyear{2020}
\copyrightyear{2020}

\history{Received: September 10, 2020; Revised: October 8, 2020; Accepted: October 19, 2020}

\usepackage{amssymb}

\DeclareMathOperator{\Tr}{Tr}

\Title{Mathematical Modeling of Japanese Encephalitis Under Aquatic Environmental Effects}

\Author{Fa\"{\i}\c{c}al Nda\"{\i}rou $^{1, 2, \dagger}$\orcidA{}, 
Iv\'{a}n Area $^{1,\dagger}$\orcidB{}
and Delfim F. M. Torres $^{2,\dagger}$*\orcidC{}}

\AuthorNames{Fa\"{\i}\c{c}al Nda\"{\i}rou, Iv\'{a}n Area and Delfim F. M. Torres}

\address{$^{1}$ \quad E. E. Aeron\'autica e do Espazo, Campus de Ourense, 
Universidade de Vigo, 32004 Ourense, Spain; area@uvigo.gal (I.A.)\\
$^{2}$ \quad Center for Research and Development in Mathematics and Applications (CIDMA),
Department of Mathematics, University of Aveiro, 3810-193 Aveiro, Portugal;
faical@ua.pt (F.N.); delfim@ua.pt (D.F.M.T.)}

\corres{Correspondence: delfim@ua.pt; Tel.: +351-234-370-668}

\firstnote{These authors contributed equally to this work.}

\abstract{We propose a mathematical model for the spread of Japanese encephalitis,
with emphasis on environmental effects on the aquatic phase of mosquitoes. 
The model is shown to be biologically well-posed and
to have a biologically and ecologically meaningful disease free equilibrium point.
Local stability is analyzed in terms of the basic reproduction number
and numerical simulations presented and discussed.}



\begin{document}
	

\section{Introduction}

Japanese encephalitis (JE) is a mosquito-borne disease transmitted 
to humans through the bite of an infected mosquito, particularly 
a \emph{Culex tritaeniorhynchus} mosquito. The mosquitoes breed 
where there is abundant water in rural agricultural areas, such 
as rice paddies, and become infected by feeding on vertebrate hosts 
(primarily pigs and wading birds) infected with the Japanese encephalitis virus. 
The virus is maintained in a cycle between those vertebrate animals and mosquitoes. 
Humans are dead-end hosts since usually they do not develop high enough 
concentrations of JE virus in their bloodstreams to infect feeding mosquitoes
\cite{BOYER20201048}.

The infection on human occasionally causes brain's inflammation 
with symptoms as headache, vomiting, fever, confusion and epileptic seizure. 
There is an estimate of about 68,000 clinical cases of occurrences with nearly 
17,000 deaths every year in Asian's countries \cite{WANG2020117080}.

The first case of Japanese encephalitis viral disease was documented in 1871 in Japan. 
But the virus itself was first isolated in 1935 and has subsequently been found across most of Asia. 
There is uncertainty on the origin's name of that virus, however phylogenetic comparisons 
with other flaviviruses suggest it evolved from an African ancestral virus, perhaps as recently 
as a few centuries ago (see \cite{solo} and references therein). Note that, despite its name, 
Japanese encephalitis is now relatively rare in Japan, as a result of a mass immunization program.

Mathematical modeling in the field of biosciences is a subject
of strong current research, see, e.g., \cite{MR3557143,MR4093642,MR4110649}.
One of the first mathematical models for the spread of JE
has been proposed and analyzed in 2009, in \cite{naresh}.
Later, in 2012, the impact of media on the spreading and control 
of JE has been carried out \cite{MR3188685}, while in 2016
several control measures to JE, such as vaccination, medicine, and insecticide,
have been investigated through optimal control and Pontryagin's maximum principle. 
The state of the art on mathematical modeling and analysis of JE,
seems to be the recent papers \cite{panja,MR3842436} of 2018.
In \cite{panja} a mathematical model on transmission of JE, 
described by a system of eight ordinary differential equations,
is proposed and studied. Main results are the basic reproduction number 
and a stability analysis around the interior equilibrium.
The authors of \cite{MR3842436} use mathematical modeling and 
likelihood-based inference techniques to try to explain the disappearance 
of JE human cases between 2006 and 2010 and its resurgence in 2011.
Here we propose a mathematical model for the spread of JE,
incorporating environmental effects on the aquatic phase 
of mosquitoes, as the primary source of reproduction. 

The manuscript is organized as follows. In Section~\ref{sec:2}, 
we introduce the mathematical model. Then, in Section~\ref{sec:3}, 
the theoretical analysis of the model is investigated: the well 
posedness of the model is proved (see Theorem~\ref{thm:wp}) and 
the meaningful disease free equilibrium and its local stability, in terms of the basic
reproduction number, analyzed in detail (see Theorem~\ref{thm:stb:dfe}). 
Section~\ref{sec:4} is then devoted to numerical simulations. 
We end with Section~\ref{sec:5} of conclusions, where we also point out
some possible directions of future research.


\section{Model Formulation}
\label{sec:2}

In our mathematical model, we shall consider environmental factors within three different host
populations: humans, mosquitoes, and vertebrate animals (pigs or wading birds) as the reservoir
host. In fact, unhygienic environmental conditions may enhance the presence and growth of vectors
(mosquitoes) populations leading to fast spread of the disease. This is due to various kinds 
of household and other wastes, discharged into the environment in residential areas of population,
and thus providing a very conducive environment for the growth of vectors \cite{ludwig, purdom}.
Since that effect could not be modeled as epidemiological compartments, we use the same scheme 
as in \cite{sinha1, sinha2} to handle that effect on the JE disease, namely \cite{naresh}
\begin{equation}
\label{model1}
\displaystyle{\frac{dE(t)}{dt}= Q_{0}+ \theta N(t) -\theta_0 E(t)},
\end{equation}
where $E$ is the cumulative density of environmental discharges conducive 
to the growth rate of mosquitoes and animals. The cumulative density 
of environmental discharges due to human activities is given by $\theta$. 
There is also a constant influx given by $Q_0$, and $\theta_0$ is the 
depletion rate coefficient of the environmental discharges. In our model, 
$N(t)$ stands for the total human population, which is considered 
a varying function of time $t$.

As for the reservoir animal populations, we consider its dynamics, 
strongly related to infected animals. Thus, the reservoir population 
constitutes a ``pool of infection'', that is a primarily source 
of infections and can be modeled by a single state variable, 
as in the framework of viruses, having free living pathogens 
in the environment (see, e.g., \cite{berge,bergeetal,code} 
and references therein for diseases like cholera, typhoid, 
or yellow fever). Therefore, we consider a single state variable, 
denoted by $I_r$, to model this reservoir pool of infection:
\begin{equation}
\label{model2}
\displaystyle{\frac{dI_r(t)}{dt}
= B\beta_{mr}\frac{I_m(t)}{N_m(t)} I_r(t) 
- \left(\mu_{1r} + \mu_{2r}I_r(t)\right) I_r(t) 
-d_r I_r(t) + \delta_0 I_r(t) E(t)},
\end{equation}
where $\displaystyle{B\beta_{mr}\frac{I_m}{N_m} I_r(t)}$ represents 
the force of infection due to interaction with mosquitoes through biting; 
$B$ is the average daily biting; $\beta_{mr}$ is the transmission coefficient 
from infected mosquitoes; $\frac{I_m}{N_m}$ is the fraction 
of infected mosquitoes; $\mu_{1r}$ the natural death rate of animals; 
$\mu_{2r}$ the density dependent death rate; $d_r$ the death rate due to the disease; 
and $\delta_0$ the per capita growth rate due to environmental discharges. 
Note that we are not interested on how the disease spread on animals.
Our main goal is to study the transmission of infections from mosquitoes to humans 
as well as the related environmental effects. 

The following assumptions are made in order to build 
the compartmental classes for mosquitoes and humans populations:
\begin{itemize}
\item we do not consider immigration of infected humans;
\item the human population is not constant (we consider a disease 
induced death rate, due to fatality, of 25\%); 
\item we assume that the coefficient of transmission of virus 
is constant and does not varies with seasons, which is reasonable 
due to the short course of the disease;
\item mosquitoes are assumed to be born susceptible. 
\end{itemize}

Three epidemiological compartments are considered for the mosquito 
population, precisely, the aquatic phase, denoted by $A_m$, 
and including eggs, larva and pupae stages; the susceptible mosquitoes, 
$S_m$; and the infected mosquitoes, $I_m$. Also, there is no resistant 
phase due to the short lifetime of mosquitoes:
\begin{equation}
\begin{cases}
\label{model3}
\displaystyle{\frac{dA_m(t)}{dt}
= \psi(1- \frac{A_m(t)}{K}) (S_m(t) + I_m(t)) 
- \left(\mu_A + \eta_A\right)A_m(t) + \delta E(t) A_m(t)},\\[3mm]
\displaystyle{\frac{dS_m(t)}{dt}= \eta_A A_m(t)- B\beta_{rm} I_r(t) S_m(t) 
- \mu_m S_m(t)},\\[3mm]
\displaystyle{\frac{dI_m}{dt}= B\beta_{rm} I_r(t) S_m(t) 
- \mu_m I_m(t)},
\end{cases}
\end{equation}
where parameter $\beta_{rm}$ represents the transmission probability 
from infected animals $I_r$ (per bite), $B$ the average daily biting, 
$\psi$ stands for the number of eggs at each deposit per capita (per day), 
$\mu_A$ is the natural mortality rate of larvae (per day), $\eta_A$ 
is the maturation rate from larvae to adult (per day), $\delta$ 
is the per capita growth rate in the level of aquatic phase 
due to conducive environmental discharge. Here $\frac{1}{\mu_m}$ denotes 
the average lifespan of adult mosquitoes (in days), 
and $K$ is the maximal capacity of larvae. We denote by $N_m$ the total 
adult mosquito populations at each instant of time $t$, being 
defined by $N_m(t)= S_m(t) + I_m(t)$ and with its dynamics satisfying 
the differential equation 
\[ 
\displaystyle{\frac{d N_m(t)}{dt}= \eta_A A_m(t) - \mu_mN_m }.
\]

The total human population, given by function $N(t)$, 
is subdivided into two mutually exclusive compartments, according 
to the disease status, namely susceptible individuals, $S$; 
and infected individuals, $I$. We do not consider a recovery 
state since there is no adequate treatment for the JE disease
and no person to person infection exists: 
\begin{equation}
\label{model4}
\begin{cases}
\displaystyle{\frac{d N(t)}{dt}
= \Lambda_h -\mu_h N(t)- d_h I(t) },\\[3mm]
\displaystyle{\frac{d S(t)}{dt} 
= \Lambda_h - \Big( B\beta_{mh} \frac{I_m(t)}{N_m(t)}\Big) S(t) 
- \mu_h S(t) + \nu_h I(t)},\\[3mm]
\displaystyle{\frac{d I(t)}{dt} 
= \Big( B\beta_{mh} \frac{I_m(t)}{N_m(t)}\Big) S(t) 
- \mu_h I(t) -\nu_h I(t) -d_h I(t)},
\end{cases}
\end{equation}
where parameter $\Lambda_h$ denotes the recruitment rate of humans, 
$\beta_{mh}$ represents the transmission probability from mosquitoes to humans, 
$\mu_h$ the natural death rate of humans, $d_h$ the disease induced death rate, 
and $\nu_h$ is the rate by which infected individuals are recovered 
and become susceptible again. The fatality rate is estimated at 25\%
of the number of infected.

Summarizing, our complete mathematical model for the JE disease is described 
by the following system of seven nonlinear ordinary differential equations:
\begin{equation}
\label{model5}
\begin{cases}
\displaystyle{\frac{dE(t)}{dt}= Q_{0}+ \theta N(t) -\theta_0 E(t)},\\[3mm]
\displaystyle{\frac{dI_r(t)}{dt}= B\beta_{mr}\frac{I_m(t)}{N_m(t)} I_r(t) 
- (\mu_{1r} + \mu_{2r}I_r(t)) I_r(t) -d_r I_r(t) + \delta_0 I_r(t) E(t)},\\[3mm]
\displaystyle{\frac{dA_m(t)}{dt}= \psi(1- \frac{A_m(t)}{K}) N_m(t) 
- (\mu_A + \eta_A)A_m(t) + \delta E(t) A_m(t)},\\[3mm]
\displaystyle{\frac{d N_m(t)}{dt}= \eta_A A_m(t) - \mu_mN_m(t) },\\[3mm]
\displaystyle{\frac{dI_m(t)}{dt}= B\beta_{rm} I_r(t) (N_m(t)-I_m(t))- \mu_m I_m(t)},\\[3mm]
\displaystyle{\frac{d N(t)}{dt}= \Lambda_h -\mu_h N(t)- d_h I(t) },\\[3mm]
\displaystyle{\frac{d I(t)}{dt} = \Big( B\beta_{mh} \frac{I_m(t)}{N_m(t)}\Big)
\Big(N(t)-I(t)\Big) - \mu_h I(t) -\nu_h I(t) -d_h I(t)},
\end{cases}
\end{equation}
where $N(t)= S(t)+I(t)$ and $N_m(t) = S_m(t) + I_m(t)$.


\section{Mathematical Analysis of the JE Model}
\label{sec:3}

We begin by proving the positivity and boundedness of solutions,
which justifies the biological well-posedness of the proposed model.

\begin{Theorem}[positivity and boundedness of solutions]
\label{thm:wp}
If the initial conditions $\left(E(0), I_r(0), A_m(0), N_m(0), I_m(0), N(0), I(0)\right)$ 
are non-negative, then the solutions $\left(E(t), I_r(t), A_m(t), N_m(t), I_m(t), 
N(t), I(t)  \right)$ of system \eqref{model5} are non-negative for all $t>0$ 
and the positive orthant $\mathbb{R}^{7}_{+}$ is positively invariant 
with respect to the flow of system \eqref{model5}. 
Furthermore, for initial conditions such that
\[
N(0)\leqslant \frac{\Lambda_h}{\mu_h} 
\quad \text{and} \quad E(0)\leqslant E^{*},
\]
one has
\[
N(t)\leqslant \frac{\Lambda_h}{\mu_h}, 
\quad E(t)\leqslant E^{*}, 
\quad I_r(t)\leqslant L, 
\quad \forall t \geqslant 0,
\]
where 
\[
E^{*}= \frac{Q_0 + \theta \frac{\Lambda_h}{\mu_h}}{\theta_0}
\quad \text{and} \quad
L= \frac{B\beta_{mr}-\mu_{1r}-d_r + \delta_0E^{*}}{\mu_{2r}}.
\]
\end{Theorem}

\begin{proof}
First of all, note that the right hand side of system \eqref{model5} 
is continuous with continuous derivatives, thus local solutions 
exist and are unique. Next, assuming that $E(0)\geqslant 0$, 
and by continuity of the right hand side of the first equation 
of system \eqref{model5}, we have that $E(t)$ remains non-negative 
on a small interval in the right hand side of $t_0=0$. Therefore, 
there exists $t_m= \sup\{t\geqslant 0: E(t)\geqslant 0\}$. 
Obviously, by definition, $t_m \geqslant 0$. To show that 
$E(t)\geqslant 0$ for all $t\geqslant 0$, we only need to prove 
that $E(t_m)>0$. Considering the first equation of system \eqref{model5}, 
that is, 
\[
\displaystyle{\frac{dE(t)}{dt}= Q_{0}+ \theta N(t) -\theta_0 E(t)},
\]
it follows that 
\[
\frac{d}{dt} \lbrace E(t)\exp(\theta_0t)\rbrace
= \Big(Q_0 + \theta N(t)\Big)\exp(\theta_0t).
\]
Hence, integrating this last equation with respect to $t$, 
from $t_0=0$ to $t_m$, we have 
\[
E(t_m)\exp(\theta_0t_m)-E(0)= \int^{t_m}_0 
\Big(Q_0 + \theta N(t)\Big)\exp(\theta_0t)dt,
\]
which yields 
\[
E(t_m)= \frac{1}{\exp(\theta_0 t_m)}\left[ E(0)
+ \int^{t_m}_0 \Big(Q_0 + \theta N(t)\Big)\exp(\theta_0t)dt \right].
\]
As a consequence, $E(t_m)> 0$ and we conclude that $E(t)>0$ for all $t>0$.
Similarly, we can prove that $I_r(t)$, $A_m(t)$, $N_m(t)$, $I_m(t)$, 
$N(t)$, and $I(t)$ are all non-negatives for all $t>0$. Moreover, 
because of the fact that $I(t)>0$ for all $t>0$, it results
from the sixth equation of system \eqref{model5} that
\[
\displaystyle{\frac{d N(t)}{dt} \leqslant \Lambda_h -\mu_h N(t) }.
\] 
Thus, applying Gronwall's inequality, we obtain 
\[
N(t)\leqslant N(0)\exp(-\mu_ht) 
+ \frac{\Lambda_h}{\mu_h}\left(1- \exp(-\mu_ht) \right).
\]
Hence, $N(t)\leqslant \frac{\Lambda_h}{\mu_h}$, 
if $N(0)\leqslant \frac{\Lambda_h}{\mu_h}$ for all $t>0$.
Further, from the first equation of system \eqref{model5} combined 
with  $N(t)\leqslant \frac{\Lambda_h}{\mu_h}$, and applying again 
Gronwall's inequality, we get $ E(t)\leqslant E^{*}$, 
whenever $E(0)\leqslant E^{*}$. From the second equation 
of system \eqref{model5}, combined with $E(t)\leqslant E^{*}$, 
we have that
\[
\frac{d I_r}{dt} \leqslant \left(A-\mu_{2r}I_r \right)I_r, 
\qquad \text{ with } \qquad A= B\beta_{mr}-\mu_{1r}
-d_r + \delta_0E^{*}.
\] 
Note that $\displaystyle{\frac{d I_r}{dt} \leqslant \left(A-\mu_{2r}I_r \right)I_r}$ 
implies $\displaystyle{\frac{1}{I_r^{2}}\frac{d I_r}{dt} \leqslant -\mu_{2r}+\frac{A}{I_r}}$ 
and, by setting $z(t)= -\displaystyle{\frac{1}{I_r}}$, we get 
\[
\frac{ d z(t)}{dt}\leqslant -\mu_{2r}-Az(t).
\]
Then we follow Gronwall's inequality to obtain that 
\[
z(t)\leqslant z(0)\exp(-At)-\frac{\mu_{2r}}{A}\left(1-\exp(-At) \right),
\]
meaning that
\[
I_r(t)\leqslant \frac{AI_r(0)}{A\exp(-At)+ \mu_{2r}I_r(0)\left(1-\exp(-At)\right)}.
\] 
Finally, $\limsup I_r(t)=\frac{A}{\mu_{2r}}$ and it follows that 
$I_r(t)\leqslant \frac{A}{\mu_{2r}}$ for all $t>0$. 
This concludes the proof.
\end{proof}

The model system \eqref{model5} admits two disease free equilibrium points (DFE),
obtained by setting the right hand side of \eqref{model5} to zero:  
a first DFE, $E_1$, given by
\[
E_1= \left(E^{*},I^{*}_r, A^{*}_m, N^{*}_m, I^{*}_m, N^{*}, I^{*}\right)
=\left( \frac{\theta \Lambda_h + Q_0\mu_h}{\theta_0\mu_h},0,0,0,0, 
\frac{\Lambda_h}{\mu_h},0 \right),
\]
corresponds to the DFE in the absence of mosquitoes population 
as well as absence of the aquatic phase, thus from a biological point 
of view this equilibrium is not interesting; a second DFE, $E_2$, 
which is the biologically and ecologically meaningful steady state 
\begin{equation}
\label{dfe}
E_2= \left(E^{*},I^{*}_r, A^{*}_m, N^{*}_m, I^{*}_m, N^{*}, I^{*}\right)
=\left( \frac{\theta \Lambda_h + Q_0\mu_h}{\theta_0\mu_h},0,\varrho,
\frac{\eta_A}{\mu_m}\varrho,0, \frac{\Lambda_h}{\mu_h},0 \right),
\end{equation}
where
\[
\varrho=\frac{K}{\psi \theta_0 \mu_h\eta_A}(\delta Q_0\mu_h \mu_m 
+ \delta \theta \Lambda_h\mu_m + \psi \eta_A \mu_h \theta_0 
-\eta_A\theta_0\mu_h \mu_m-\theta_0\mu_h \mu_m \mu_A),
\]
which can be rewritten as
\begin{align*}
\varrho &=\frac{K}{\psi \theta_0 \mu_h\eta_A}\left(\delta\mu_m 
(\theta\Lambda_h + Q_0\mu_h) + \psi \eta_A \mu_h \theta_0 
-\eta_A\theta_0\mu_h \mu_m-\theta_0\mu_h \mu_m \mu_A \right)\\
&= \frac{K}{\psi}\left( \frac{\delta \mu_m}{\eta_A}E^{*} 
+ \psi -\mu_m -\frac{\mu_m\mu_A}{\eta_A}\right).
\end{align*}
Therefore,
\begin{equation}
\label{nmstar}
N^{*}_m= \frac{\eta_A}{\mu_m}\varrho= \frac{K}{\psi}\left(\delta E^{*} 
+ \frac{\eta_A \psi}{\mu_m}-\eta_A -\mu_A \right).
\end{equation}

The equilibrium $E_2$ considers interaction with mosquito populations 
and, with that, aquatic phase as initial source of mosquito's reproduction.

We compute the basic reproduction number using the next generator matrix 
method as described in \cite{van}. In doing so, we consider the following 
set of vectors:
\begin{equation*}
\mathcal{F} =
\left(
\begin{array}{c}
\delta_0I_r E + B\beta_{mr}\frac{I_m}{N_m} I_r \\[3mm]
B\beta_{rm}I_r(N_m -I_m)\\[3mm]
\displaystyle{\frac{B\beta_{mh}I_m(N-I)}{N}}
\end{array}
\right)
\end{equation*}
and
\begin{equation*}
\mathcal{V} =
\left(
\begin{array}{c}
(\mu_{1r} + \mu_{2r}I_r) I_r + d_r I_r \\[3mm]
\mu_mI_m\\[3mm]
(\mu_h + \nu_h + d_h)I
\end{array}
\right).
\end{equation*}
Then, we compute the Jacobian matrix associated to $\mathcal{F}$ 
and $\mathcal{V}$ at the DFE, $E_2$, that is, 
\begin{equation*}
J_{\mathcal{F}}
= \left[ \begin{array}{ccc}
\delta_0E^{*} & 0 & 0\\
B\beta_{rm}N^{*}_m& 0 & 0\\
0& B \beta_{mh}& 0
\end{array}
\right],
\quad
J_{\mathcal{V}}= \left[ 
\begin{array}{ccc}
d_r + \mu_{1r}& 0 & 0\\
0& \mu_m & 0\\
0& 0& \mu_h + \nu_h + d_h 
\end{array}
\right].
\end{equation*}
The basic reproduction number $R_0$ is obtained as the spectral 
radius of the matrix $J_{\mathcal{F}}\times (J_{\mathcal{V}})^{-1}$ 
at the disease free equilibrium $E_2$, being given by 
\begin{equation}
\label{eq:R0}
R_0= \frac{\delta_0 E^{*}}{d_r + \mu_{1r}}
= \left(\delta_0\frac{\theta \Lambda_h + Q_0\mu_h}{\theta_0\mu_h}\right)
\times \frac{1}{ d_r + \mu_{1r}}.
\end{equation}

The local stability of the disease free equilibrium (DFE) 
can be studied through an eigenvalue problem 
of the linearized system associated to (\ref{model5}) at the DFE $E_2$. 
The DFE point is locally asymptotically stable if all the eigenvalues, 
of the matrix representing the linearized system associated to (\ref{model5}) at the DFE $E_2$, 
have negative real parts \cite{steven}. The aforementioned matrix is given by
\begin{equation*}
M=\left[ 
\begin{array}{ccccccc}
-\theta_0 & 0&0&0&0&0&0\\
0& M_{22}& 0&0&0&0&0\\
\delta A^{*}_m & 0& M_{33}& \psi \Big( 1- \frac{A^{*}_m}{K}\Big) & 0& 0&0\\
0& 0& \eta_A & -\mu_m & 0 &0&0 \\
0& B\beta_{rm}N^{*}_m & 0 &0& -\mu_m &0& 0\\
0&0&0&0&0& -\mu_h & -d_h\\
0& 0& 0& 0& B\beta_{mh}& 0& -\mu_h - \nu_h -d_h
\end{array}
\right],
\end{equation*}
where $M_{22}= \delta_0 E^{*} - d_r-\mu_{1r}$ 
and $M_{33}= -\frac{\psi N^{*}_m}{K}-\mu_A -\eta_A + \delta E^{*}
= -\frac{\eta_A \psi}{\mu_m}$, by using \eqref{nmstar}. 
The eigenvalues of this matrix are 
\begin{gather*}
\lambda_1 = -\theta_0, 
\quad \lambda_2= \delta_0 E^{*} - d_r-\mu_{1r}= (d_r+\mu_{1r})( R_0 -1), \\ 
\lambda_3= -\mu_m, \quad \lambda_4= -\mu_h, \quad \lambda_5= -\mu_h - \nu_h -d_h
\end{gather*}
and the other two remaining eigenvalues are of the following square matrix: 
\begin{equation*}
J= \left[ 
\begin{array}{cc}
-\frac{\eta_A \psi}{\mu_m}& \psi\Big(1- \frac{A^{*}_m}{K} \Big)\\
\eta_A & -\mu_m
\end{array} \right].
\end{equation*}
Since the trace of this matrix, 
$\Tr J = -\displaystyle{\frac{\eta_A \psi}{\mu_m}- \mu_m}$, 
is negative, and its determinant
\begin{equation*}
\det J= \displaystyle{\eta_A\psi- \eta_A \psi\Big(1- \frac{A^{*}_m}{K} \Big)
= \frac{A^{*}_m}{K}}
\end{equation*}
positive, it follows that these two eigenvalues are both negative. 
In conclusion, we have just proved the following result.

\begin{Theorem}[local stability of the biologically and ecologically meaningful
disease free equilibrium]
\label{thm:stb:dfe}
The disease free equilibrium $E_2$ with aquatic phase 
and in the presence of non-infected mosquitoes is locally 
asymptotically stable if $R_0 < 1$ and unstable if $R_0 > 1$, 
where $R_0$ is given by \eqref{eq:R0}. 
\end{Theorem}


\section{Numerical Simulations}
\label{sec:4}

\begin{figure}[!htb]
\centering 
\includegraphics[scale=0.77]{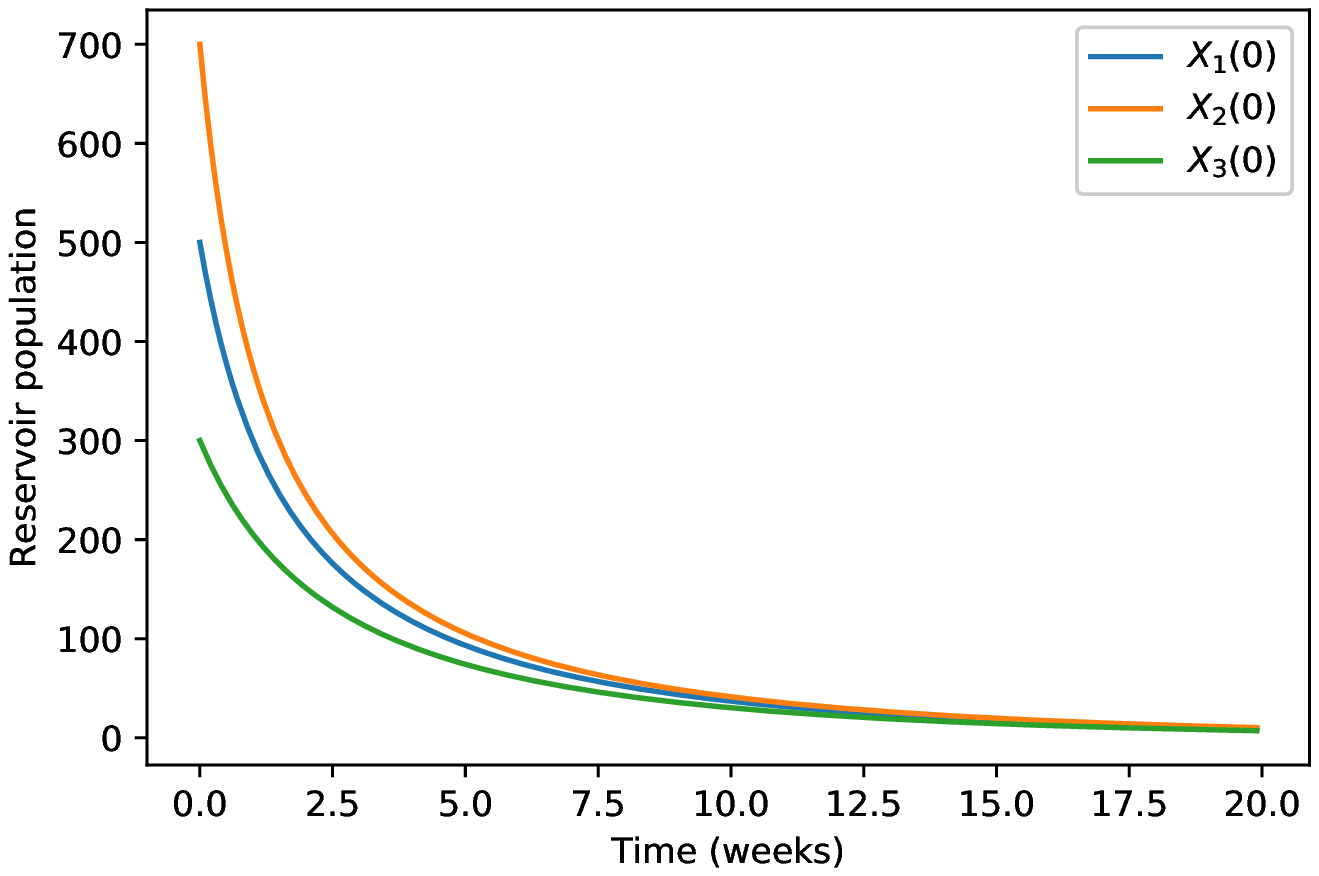}
\caption{The solution of model \eqref{model5} tends to the disease free equilibrium. 
In this figure, we show the evolution of the infected animals population 
for different initial conditions.}
\label{Inf:reservoir}
\end{figure}

\begin{figure}[!htb]
\centering 
\includegraphics[scale=0.77]{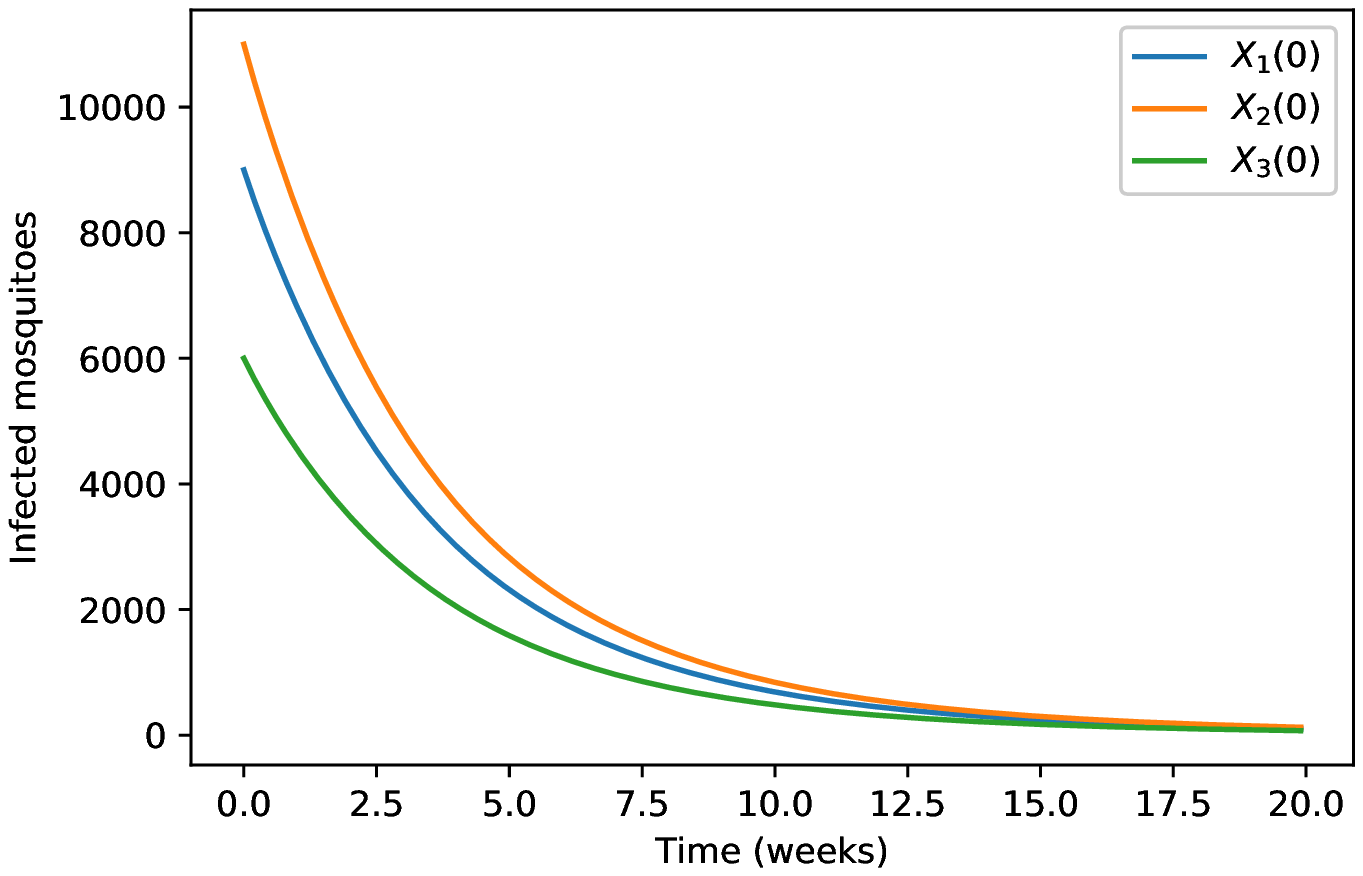}
\caption{The solution of model \eqref{model5} tends to the disease free equilibrium. 
In this figure, we show the evolution of the infected mosquitoes population 
for different initial conditions.}
\label{Inf:mosquioes}
\end{figure}

\begin{figure}[!htb]
\centering 
\includegraphics[scale=0.77]{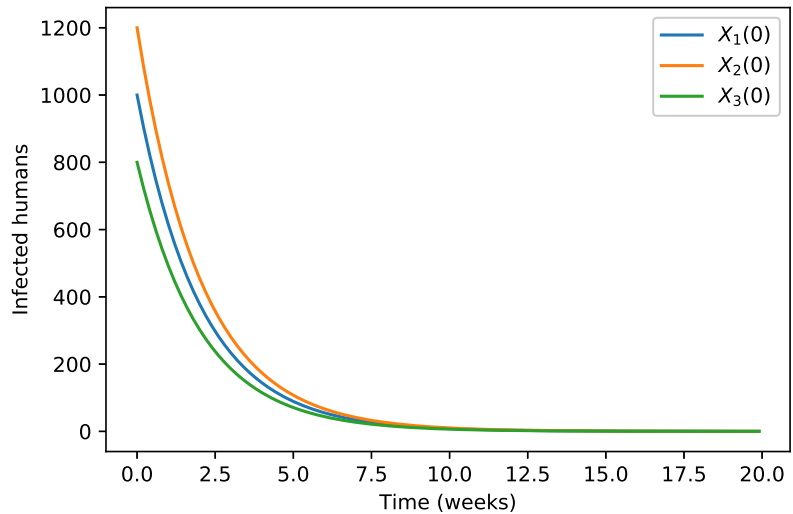}
\caption{The solution of model \eqref{model5} tends to the disease free equilibrium. 
In this figure, we show the evolution of the infected humans population 
for different initial conditions.}
\label{Inf:humans}
\end{figure}

In this section, we illustrate stability and convergence of the
solutions of the differential system \eqref{model5} 
to the disease free equilibrium \eqref{dfe} for different values 
of initial conditions considered in Table~\ref{tab:initialcondition} 
(see Figures~\ref{Inf:reservoir}, \ref{Inf:mosquioes}, and \ref{Inf:humans} 
for the corresponding infected populations in model \eqref{model5}). We perform 
numerical simulations to solve the model system \eqref{model5} by using the \textsf{Python} 
programming language, precisely the freely available routine 
\texttt{integrate.odeint} of library \textsf{SciPy}.
The following values of the parameters, borrowed from \cite{naresh,panja}, 
are considered:
\begin{gather*}
Q_0= 50, \quad \theta =0.0002, \quad \theta_0 = 0.0001, 
\quad \beta_{mr}= 0.0001, \quad \mu_{1r}= 0.1, \quad dr=1/15, \\ 
\delta_0= 0.000001, \quad \Lambda_h= 150, \quad \mu_h= 1/65, \quad d_h= 1/45, 
\quad \nu_h = 0.45,\quad  \beta_{mh}=0.0003, \\ 
\psi = 0.6, \quad K=1000, \quad \delta= 0.0001, 
\quad \mu_m= 0.3, \quad \beta_{rm}= 0.00021.
\end{gather*}
Moreover, the remaining parameters were estimated as follows:
\begin{gather*}
\mu_{2r}= 0.001, \quad \eta_{A}=0.5 , \quad \mu_{A}= 0.25, \quad B=1.
\end{gather*}

The value of the DFE, $E_2$ is computed as below:
\begin{equation*}
E_2= \left(E^{*},I^{*}_r, A^{*}_m, N^{*}_m, I^{*}_m, N^{*}, I^{*}\right)
=\left(122.959, \, \,0, \, \,262.296, \,\,437.160, \, \,0, \, \,9750, \, \,0 \right).
\end{equation*}
The matrices $J_{\mathcal{F}}$ and $J_{\mathcal{V}}$ are obtained as follows
\begin{equation*}
J_{\mathcal{F}}
= \left[ \begin{array}{ccc}
0.000123 & 0 & 0\\
0.0918& 0 & 0\\
0& 0.0003& 0
\end{array}
\right],
\quad
J_{\mathcal{V}}= \left[ 
\begin{array}{ccc}
0.167& 0 & 0\\
0& 0.3 & 0\\
0& 0& 0.488
\end{array}
\right],
\end{equation*}
which leads to the value of $R_0= 0.000738$.

Furthermore, we have that the matrix $M$ is equal to 
\begin{equation*}
M=\left[ 
\begin{array}{ccccccc}
-0.0001 & 0&0&0&0&0&0\\
0& -0.167& 0&0&0&0&0\\
0.0262 & 0& -1& 0.443 & 0& 0&0\\
0& 0& 0.5 & -0.3 & 0 &0&0 \\
0& 0.0918 & 0 &0& -0.3 &0& 0\\
0&0&0&0&0& -0.0154 & -0.0222\\
0& 0& 0& 0& 0.0003& 0& -0.487
\end{array}
\right],
\end{equation*}
and its eigenvalues are $$
-1.236, \, \, -0.0636,\, \,-0.0001,\, \,-0.0154, \, \,-0.488, \, \,-0.3, \, \,-0.167$$
all negatives in accordance with Theorem \ref{thm:stb:dfe}, since $R_0<1$. 

The initial conditions were considered as in Table~\ref{tab:initialcondition} 
and the evolution of the three infected populations are strictly decreasing curves 
with all of them converging to the disease free equilibrium (Figures~\ref{Inf:reservoir}--\ref{Inf:humans}) 
for this specific parameter values.
\begin{table}[!htb]
\centering
\caption{Initial conditions considered}
\label{tab:initialcondition}
\begin{tabular}{|c|c|c|c|c|c|c|c|} \hline
&$E(0)$ & $I_r(0)$ & $A_m(0)$ &$ N_m(0)$ & $I_m(0)$& $N(0)$ & $I(0)$ \\[1mm] \hline
$X_1(0)$&$40000$ & $500$& $12000$ &$ 10000$ & $9000$& $7000$ & $1000$ \\[1mm] \hline
$X_2(0)$&$45000$ & $700$& $15000$ &$ 12000$ & $11000$& $10000$ & $12000$ \\[1mm] \hline
$X_3(0)$&$35000$ & $300$& $10000$ &$ 7000$ & $6000$& $5000$ & $800$ \\[1mm] \hline
\end{tabular}
\end{table}

\begin{figure}[!htb]
\centering 
\includegraphics[scale=0.77]{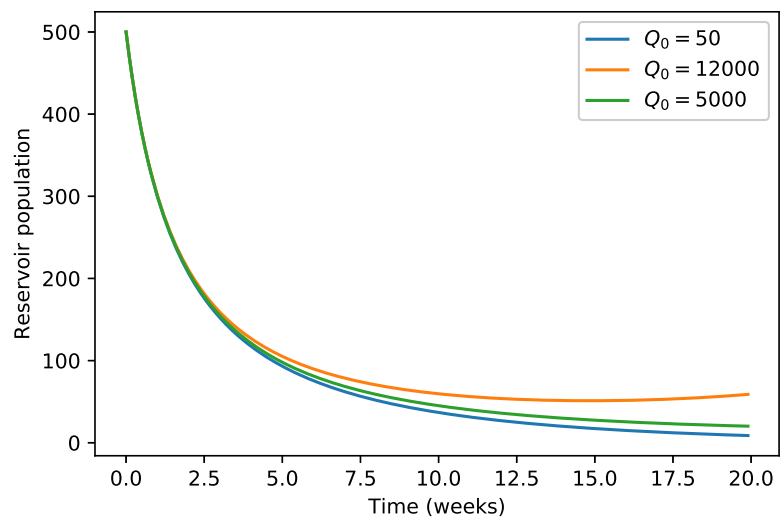}
\caption{Variation of animals population with respect to $Q_0$.}
\label{Inf:reservoir_Q0}
\end{figure}

\begin{figure}[!htb]
\centering 
\includegraphics[scale=0.77]{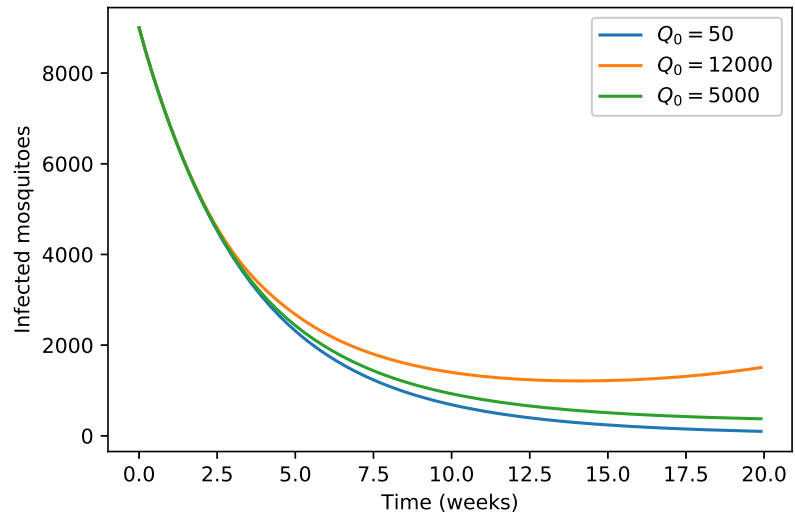}
\caption{Variation of infected mosquitoes population with respect to $Q_0$.}
\label{Inf:mosquioes_Q0}
\end{figure}

Our numerical simulations show that the evolution of the three infected populations 
are strictly decreasing curves and, all of them, converge to the disease free equilibrium
(Figures~\ref{Inf:reservoir}--\ref{Inf:humans}). This means that our Japanese Encephalitis
model \eqref{model5} describes a situation of an epidemic disease through an interesting 
environmental effect on the source of reproduction of mosquitoes, namely the aquatic phase 
of mosquitoes, which includes eggs, larva, and pupa stages. Furthermore, in 
Figures~\ref{Inf:reservoir_Q0}--\ref{Inf:mosquioes_Q0} the variation of the evolution 
of the infected animals population and infected mosquitoes population is shown, 
respectively with respect to different values in the level of environmental discharge 
due to constant influx $(Q_0)$. It is found that with the decrease in the level of environmental 
discharge due to constant influx $(Q_0)$, the infected animals population and infected 
mosquitoes population decrease and approach the disease free equilibrium state.


\begin{figure}[!htb]
\centering 
\includegraphics[scale=0.77]{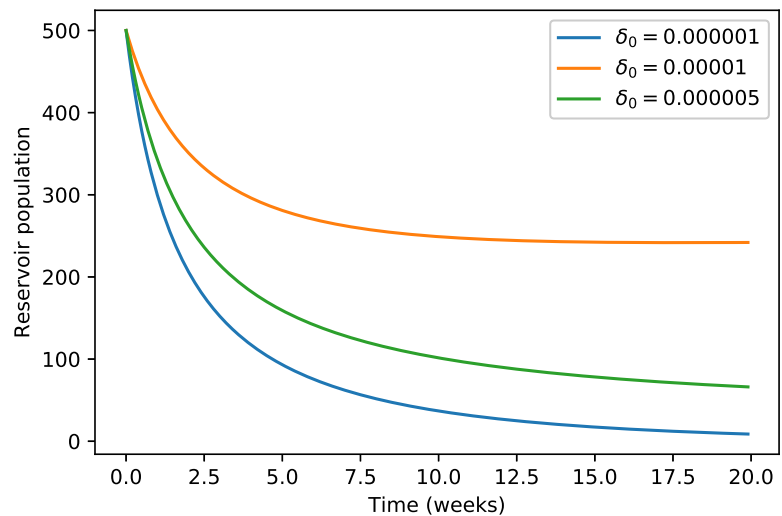}
\caption{Variation of animals population with respect to $\delta_0$.}
\label{Inf:reservoir_delta0}
\end{figure}

\begin{figure}[!htb]
\centering 
\includegraphics[scale=0.77]{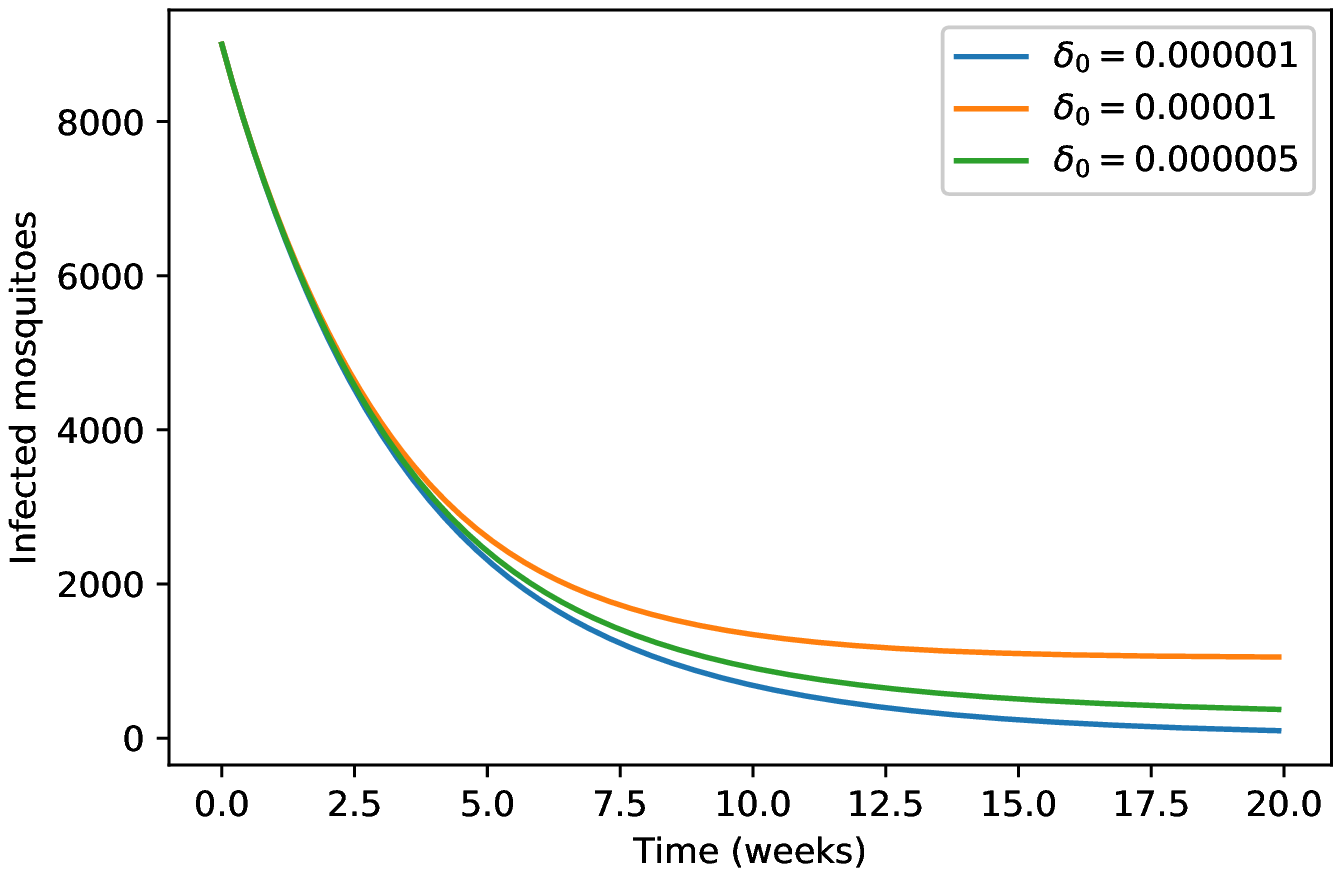}
\caption{Variation of infected mosquitoes population with respect to $\delta_0$.}
\label{Inf:mosquioes_delta0}
\end{figure}

We observe in Figures~\ref{Inf:reservoir_delta0}--\ref{Inf:mosquioes_delta0} 
that the decrease of the per capita growth rate $\delta_0$ of animals due 
to environmental discharges, results in the decrease of infected animals population 
as well as for infected mosquitoes population. 


\begin{figure}[!htb]
\centering 
\includegraphics[scale=0.77]{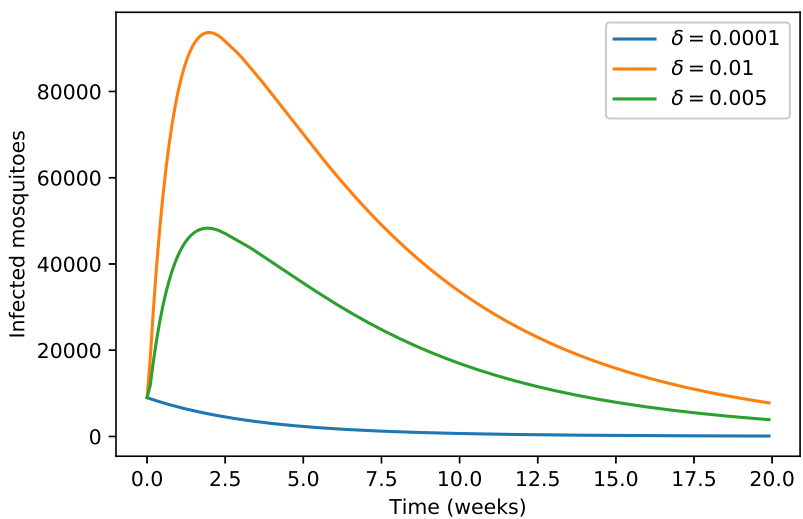}
\caption{Variation of infected mosquitoes population with respect to $\delta$.}
\label{Inf:mosquioes_delta}
\end{figure}

The role of conducive environmental discharge $\delta$ on the infected mosquitoes 
population is shown in Figure~\ref{Inf:mosquioes_delta}. We found that when the 
value of $\delta$ is smaller than $0.0001$, then there is a strict decrease 
in the number of infected mosquitoes population. However, when $\delta$ becomes larger, 
the infected mosquitoes population increases up to a certain optimum value 
and then decreases to the disease free equilibrium state.

\section{Conclusions} 
\label{sec:5}

In \cite{naresh}, a Japanese Encephalitis model is studied. 
Their results show persistence of disease in the population, that is, 
an endemic situation. In contrast, our obtained results highlight 
the importance of considering environmental effects on the aquatic phase 
of mosquitoes, as the primary source of reproduction of mosquitoes.
This is not considered in \cite{naresh}, where the environmental 
effect is acting on the mature susceptible mosquitoes populations. 
Here we have shown that the basic reproduction number is a linear
dependent function with respect to the equilibrium state of 
the cumulative density of environmental discharges, 
conducive to the growth rate of mosquitoes and animals. 
All our computational experiments were carried out using 
the free and open-source scientific computing \textsf{Python} 
library \textsf{SciPy}. To make our results reproducible, 
we provide the main computer code in Appendix~\ref{app:A}.
As future work, it would be interesting to validate the model 
with real data; and take into account possible control measures, e.g.,
vaccination of the population and vector or environmental controls.


\appendix

\section{\textsf{Python} code for Figures~1, 2 and 3}
\label{app:A}

{\small
\begin{verbatim}
"""" Numerical simulations for japaneese encephalitis disease """

# import modules for solving 
import scipy
import scipy.integrate
import numpy as np

# import module for plotting
import pylab as pl

# System with substittutions 
#E=X[0], I_r = X[1]; A_m=X[2]; N_m=X[3]; I_m=[4]; N=X[5]; I=X[6].
def JEmodel(X, t, Q0, theta, theta0, betamr, mu1r, mu2r,  dr, delta0, psi, K, muA, nuA, 
    delta, mum, B, betarm, Lambdah, muh, nuh, dh, betamh ):
    	z1= Q0 + theta*X[5] - theta0*X[0]
	z2=betamr*X[1]*X[4]/X[3] - (mu1r +mu2r*X[1] + dr)*X[1] + delta0*X[1]*X[0] 
    	z3= psi*(1- X[2]/K)*X[3] - (muA + nuA)*X[2] + delta*X[0]*X[2]
    	z4= nuA*X[2]-mum*X[3]
    	z5= B*betarm*X[1]*(X[3]-X[4])-mum*X[4]
    	z6= Lambdah - muh*X[5]-dh*X[6]
    	z7= (B*betamh*X[4]/X[3])*(X[5]-X[6])-nuh*X[6] - muh*X[6] - dh*X[6]
	return (z1, z2, z3, z4, z5, z6,z7)

if __name__== "__main__":
	
	X0= [40000, 500, 12000, 10000, 9000, 7000, 1000];
	X1= [45000, 700, 15000, 12000, 11000, 10000, 1200];
	X2= [35000, 300, 10000, 7000, 6000, 5000, 800];
	t = np.arange(0, 20, 0.1) 
	
	Q0= 50
	theta=0.01
	theta0=0.0001
	betamr= 0.0001
	mu1r=0.1
	dr=1/15.0
	delta0=0.000001
	psi=0.6
	K=1000
	muA=0.25
	nuA=0.5
	delta=0.0001
	mum=0.3
	B=1; mu2r= 0.001
	betarm=0.00021
	Lambdah=150
	muh=1.0/65
	dh=1.0/45
	nuh=0.45
	betamh=0.0003
	r=scipy.integrate.odeint(JEmodel, X0, t, args=(Q0, theta, theta0, betamr, mu1r, mu2r,  
    dr, delta0, psi, K, muA, nuA, delta, mum, B, betarm, Lambdah, muh, dh, nuh,  betamh))

	r1=scipy.integrate.odeint(JEmodel, X1, t, args=(Q0, theta, theta0, betamr, mu1r, mu2r,  
    dr, delta0, psi, K, muA, nuA, delta, mum, B, betarm, Lambdah, muh, dh, nuh,  betamh))

	r2=scipy.integrate.odeint(JEmodel, X2, t, args=(Q0, theta, theta0, betamr, mu1r, mu2r,  
    dr, delta0, psi, K, muA, nuA, delta, mum, B, betarm, Lambdah, muh, dh, nuh,  betamh))
	
	pl.plot(t,r[:,1], t,r1[:,1], t,r2[:,1])
	pl.legend(['$X_1(0)$', '$X_2(0)$', '$X_3(0)$'],loc='upper right')
	pl.xlabel('Time (weeks)')
	pl.ylabel('Reservoir population')
	#pl.title('Japaneese model')
	pl.savefig('reservoir.eps')
	pl.show();  
		
	pl.plot(t,r[:,4],  t,r1[:,4],t,r2[:,4])	
	pl.xlabel('Time (weeks)')
	pl.ylabel('Infected mosquitoes')
	pl.legend(['$X_1(0)$', '$X_2(0)$', '$X_3(0)$'],loc='upper right')
	pl.savefig('mosquitoes.eps')
	pl.show();  

	pl.plot(t,r[:,6], t,r1[:,6],t,r2[:,6])
	pl.xlabel('Time (weeks)')
	pl.ylabel('Infected humans')
	pl.legend(['$X_1(0)$', '$X_2(0)$', '$X_3(0)$'],loc='upper right')
	pl.savefig('infected_human.eps')
	pl.show()
\end{verbatim}
}


\authorcontributions{The authors equally contributed 
to this paper, read and approved the final manuscript:
Formal analysis, Fa\"{\i}\c{c}al Nda\"{\i}rou, Iv\'{a}n Area and Delfim F. M. Torres; 
Investigation, Fa\"{\i}\c{c}al Nda\"{\i}rou, Iv\'{a}n Area and Delfim F. M. Torres; 
Writing -- original draft, Fa\"{\i}\c{c}al Nda\"{\i}rou, Iv\'{a}n Area and Delfim F. M. Torres; 
Writing -- review \& editing, Fa\"{\i}\c{c}al Nda\"{\i}rou, Iv\'{a}n Area and Delfim F. M. Torres.}

\funding{This research was partially funded by the Portuguese 
Foundation for Science and Technology (FCT) through CIDMA,
grant number UIDB/04106/2020 (F.N. and D.F.M.T.);
and by the Agencia Estatal de Investigaci\'on (AEI) of Spain under Grant MTM2016-75140-P, 
cofinanced by the European Community fund FEDER (I.A.).
F.N. was also supported by FCT through the PhD fellowship PD/BD/150273/2019.} 

\acknowledgments{The authors are grateful to four anonymous reviewers 
for several pertinent questions and comments.}

\conflictsofinterest{The authors declare no conflict of interest.} 


\reftitle{References}


\end{document}